\documentclass[reprint,amsmath,amssymb,aps,superscriptaddress,fleqn]{revtex4-2}
\usepackage{amsmath}
\usepackage{hyperref}
\usepackage{graphicx}
\usepackage{adjustbox}
\usepackage{bm}

\begin{document}

\title{Microscopic nucleus-nucleus optical potentials from nuclear matter \\ with uncertainty analysis from chiral forces}

\author{T. R. Whitehead}
\affiliation{Facility for Rare Isotope Beams, Michigan State University, East Lansing, Michigan 48824, USA}

\begin{abstract}
Nucleus-nucleus optical potentials are constructed from an energy density functional approach first outlined by Brueckner et al. The interaction term of the energy density functional comes from the complex nucleon self-energy computed in nuclear matter with two- and three-body chiral nuclear forces. Nuclear density distributions are calculated from Skyrme functionals constrained to the equations of state calculated from the same chiral forces used for the self-energy. Predictions for elastic scattering cross sections and fusion cross sections are compared to experimental data. Very good agreement is found with experiment for elastic scattering of heavier nucleus-nucleus systems at energies in the range of $20 < E < 90$ MeV/N, while accurate descriptions of lighter and lower-energy systems may require the inclusion of collective excitations.
\end{abstract}

\maketitle

\emph{Introduction} - As the field of low-energy nuclear theory increasingly turns its focus to exotic nuclei, theoretical predictions with uncertainty quantification will be essential for experimental efforts at radioactive beam facilities. In past studies of nuclei near stability, phenomenological reaction models tuned to experimental data were considered sufficient. However, as modern experimental capabilities allow for studies of rare isotopes and as advances in nuclear many-body theory and our understanding of the nuclear force enable predictive microscopic calculations of exotic nuclei, it is imperative that nuclear reaction models be further developed to accommodate the needs of experimental efforts.

Optical potentials play a central role in the theoretical modeling of nuclear reactions and the interpretation of experimental data. There has been much recent development in the area of microscopic nucleon-nucleus optical potentials \cite{Whitehead21,Vorrabi22,Baker21,Idini19,Rotureau18}. However, most reaction experiments in the rare isotope era will involve nucleus-nucleus interactions and therefore require nucleus-nucleus optical potentials. The study of reaction channels such as knockout, transfer, charge-exchange, and fusion could benefit substantially from the implementation of microscopic nucleus-nucleus optical potentials with quantified uncertainties. In contrast to nucleon-nucleus optical potentials, less attention has been paid to the development of microscopic nucleus-nucleus optical potentials, with a few notable exceptions that focus on light systems. These works follow the double-folding approach \cite{SATCHLER79} where either a free-space $N\!N$ interaction \cite{DURANT18,Durant20,Durant22} or a g-matrix interaction \cite{Minomo14,Minomo16} is folded with nuclear density distributions. When a free-space $N\!N$ interaction is employed, the double folding approach yields a real interaction which must be supplemented by an imaginary term to account for absorption. Typically this is done by assuming the real and imaginary radial dependences to be equal and adjusting the strength of the imaginary term to reproduce experimental data. Recently in Ref. \cite{Durant20}, a more theoretically rigorous approach is taken by utilizing the dispersion relation to derive the strength of the imaginary term from the real part. Another important distinction of the work by Durant et al. \cite{DURANT18,Durant20,Durant22} amongst the double folding approaches is the estimation of the theoretical uncertainty by varying the radial cutoff of a local chiral $N\!N$ interaction. Aside from these works, the quantification of theoretical uncertainties in the calculation of nucleus-nucleus optical potentials has been largely neglected. In Refs. \cite{Minomo14,Minomo16} a double folding approach is carried out using the Melbourne g-matrix interaction modified to approximately account for effects from three-body forces. By utilizing the g-matrix instead of a free-space interaction, in-medium effects are accounted for. In Ref. \cite{Minomo16} the authors show the importance of including collective excitations in a coupled-channels framework. They achieve substantially better agreement with data in the coupled-channels calculations, however discrepancies still persist at large scattering angles.

An alternative method for constructing the nucleus-nucleus optical potential is through an energy density functional based on the nuclear matter single-nucleon potential. Brueckner et al. first outlined how energy density functionals may be used to construct real nucleus-nucleus interactions in Refs. \cite{Brueckner68,Brueckner68_2}. This energy density functional approach may be generalized to include the imaginary term of the nucleus-nucleus optical potential on equal footing to the real term \cite{GUPTA78,Behera80,Muller80,FAESSLER81,IZUMOTO80,IZUMOTO81,OHTSUKA87,TREFZ84,TREFZ85} in contrast to the double folding approach when carried out with an $N\!N$ interaction. Another advantage of utilizing many-body calculations in nuclear matter is that Pauli blocking effects and correlations are included.

The present work constructs a nucleus-nucleus optical potential starting from many-body perturbation theory calculations of the self-energy in nuclear matter, which are also the basis of the WLH global nucleon-nucleus optical potential \cite{Whitehead21}. The results of the current work may be combined with the WLH model for a consistent microscopic treatment of the effective interactions in few-body reaction models. Two-body nuclear forces from chiral effective field theory (EFT) calculated to N$^3$LO with three-body forces at N$^2$LO \cite{Entem03,Coraggio07,coraggio14,Coraggio13,Marji13} are used in the calculation of the nuclear matter self-energy. To estimate the theoretical uncertainty from the nuclear force in predictions of nuclear reaction observables, three chiral interactions with momentum space cutoffs in the range of $\Lambda$= 414-500 MeV are employed. Results are benchmarked by constructing optical potentials for a variety of systems where elastic scattering data are available for comparison. Fusion reactions are also used as a benchmark for the nucleus-nucleus interactions, as they are important in many contexts ranging from supernova nucleosynthesis to experiments at rare isotope beam facilities. Fusion cross sections are calculated for a set of systems and compared to recent studies also based on Brueckner's approach \cite{Umar21,Simenel17}.

\emph{Formalism} - The present work follows Brueckner \cite{Brueckner68} in constructing a nucleus-nucleus ($A_1$+$A_2$) interaction from the nuclear matter self-energy through an energy density functional. An imaginary term is also included as in Ref. \cite{GUPTA78}. The $A_1$+$A_2$ interaction is obtained by integrating the energy density of the combined system minus the energy densities of each isolated nucleus over space for a given separation distance $R$.

\begin{align}
U(R,E) = \int [H(\rho_T,E) - H(\rho_1,E) - H(\rho_2,E) ] d^3\bar{r}
\end{align}
In the first term of the integrand, the density $\rho$ and isospin asymmetry $\delta$ are taken to be the sum of each nucleus where the total density is $\rho_T=\rho_1+\rho_2$ and total isospin asymmetry is $\delta_T=\frac{\rho_1^n+\rho_2^n-\rho_1^p-\rho_2^p}{\rho_1+\rho_2}$. This so-called frozen density approximation assumes the two nuclear densities simply add together, neglecting Pauli blocking effects between nucleons in different nuclei. Improvements on this assumption will be the focus of future works. The density and isospin asymmetry in the second and third terms of the integrand in Eq. (1) are of each isolated nucleus. The form of the energy density is given by
\begin{align}
\begin{split}
H\Big(\rho\big(\bar{r},\bar{R}\big),E\Big)& = \frac{3}{10M_N} \big(3\pi^2\big)^{2/3} \rho\big(\bar{r},\bar{R}\big)^{5/3} \\ 
&+ \frac{1}{72M_N} \frac{\Big(\nabla \rho\big(\bar{r},\bar{R}\big) \Big)^2}{\rho\big(\bar{r},\bar{R}\big)} \\
&+ \rho\big(\bar{r},\bar{R}\big) U\Big(\rho\big(\bar{r},\bar{R}\big) , \delta\big(\bar{r},\bar{R}\big), E\Big).
\end{split}
\end{align}
 In this work, the single-nucleon potential $U$ is the complex and energy dependent nuclear matter self-energy $U(\rho,\delta,E)=V(\rho,\delta,E)+iW(\rho,\delta,E)$. The self-energy is taken to be the average between the proton and neutron self-energies weighted by the isospin asymmetry at a given point in the $A_1$+$A_2$ system:

\begin{figure}
\includegraphics[scale=0.31]{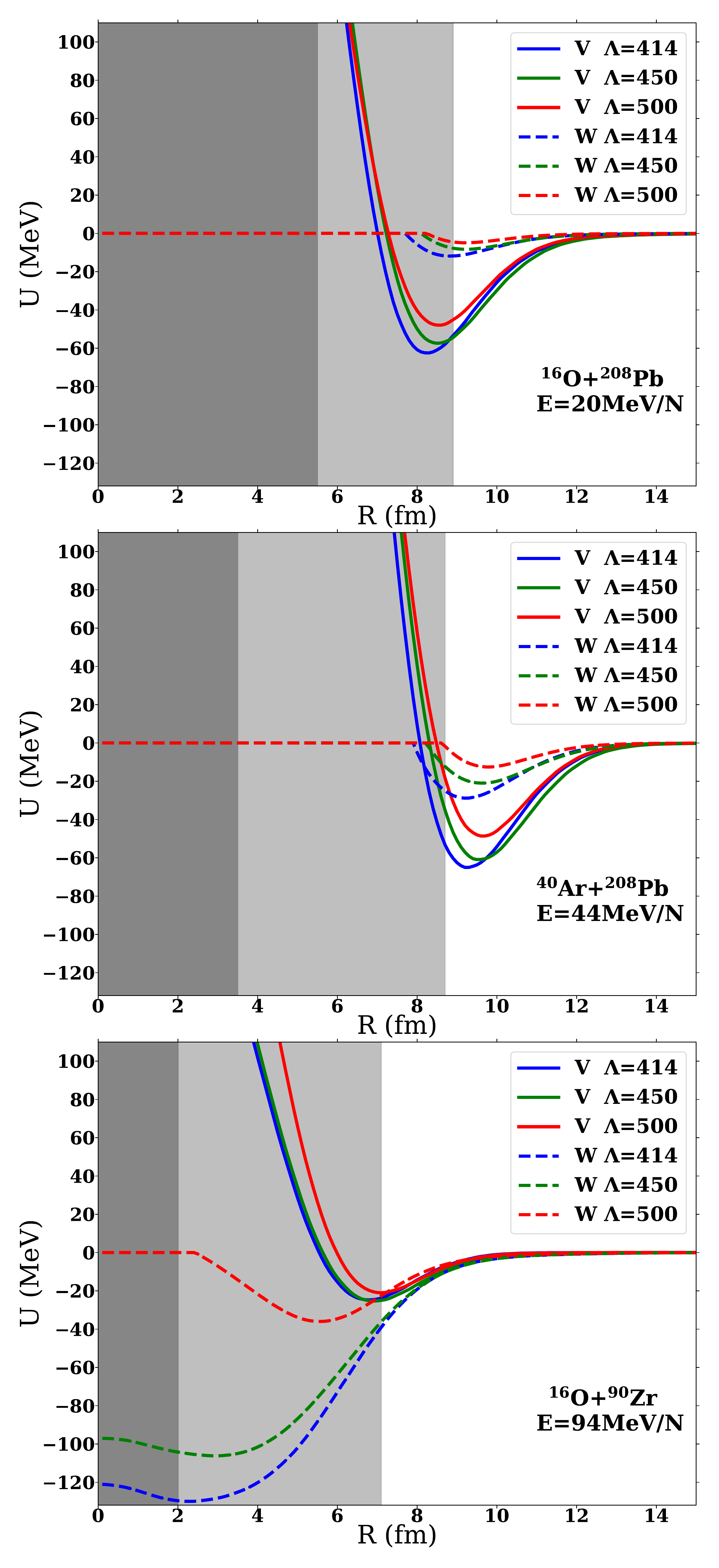}
\caption{The real ($V$) and imaginary ($W$) terms of a selection of optical potentials as a function of the separation distance between the nuclei. Results from N$^3$LO chiral interactions with momentum space cutoffs of $\Lambda$ = 414, 450, 500 MeV are shown in blue, green, and red. The light grey region represents where the frozen density approximation yields densities beyond nuclear saturation. The dark grey region represents the region of the potential that does not affect elastic scattering cross sections for the given case.\label{Potential}}
\end{figure}

\begin{figure*}
\begin{center}
\includegraphics[scale=0.36]{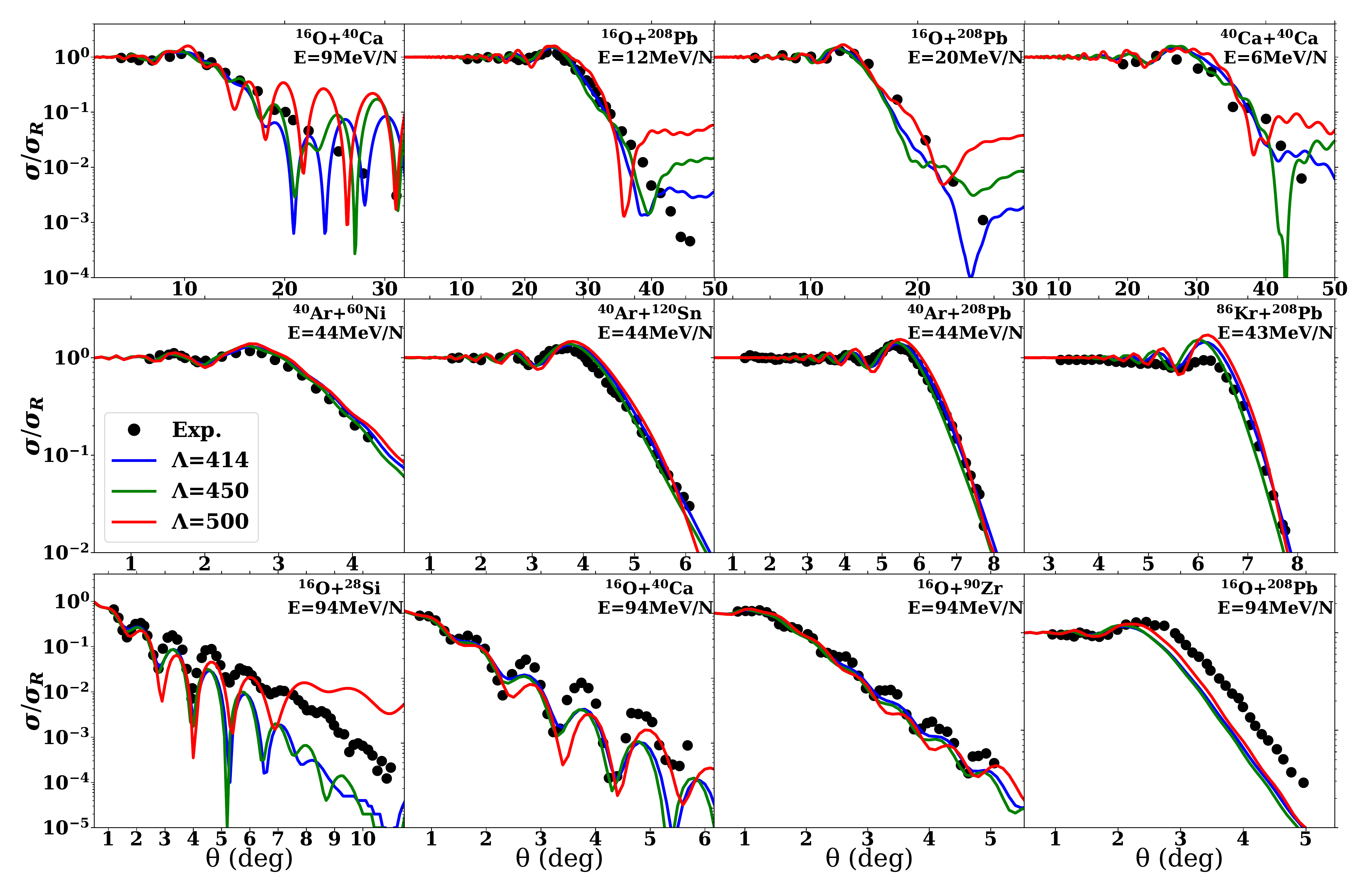}
\caption{Elastic scattering cross sections for a selection of medium- and heavy-mass nucleus-nucleus systems at varied energies. Results from N$^3$LO chiral interactions with momentum space cutoffs of $\Lambda$ = 414, 450, 500 MeV are shown in blue, green, and red. Experimental data from Refs. \cite{ALAMANOS84,Roussel88,ROUSSEL88_2,Doubre77,Vigdor79,Pieper78} are shown as black dots. \label{CrossSection}}
\end{center}
\end{figure*} 

\begin{equation}
U  = \frac{ \big(1+\delta \big) U_n + \big(1-\delta \big) U_p}{2}.
\end{equation}
Similarly, the energy $E$ is taken to be the average energy of each nucleus in the lab frame $E_1,E_2$ weighted by their densities:

\begin{equation}
E  = \frac{ \rho_1 E_1 + \rho_2 E_2}{\rho_1+\rho_2}.
\end{equation}
The first term of Eq. (2) is the kinetic energy and the second term is its gradient correction in the Thomas-Fermi approximation. The third term represents the potential energy. For a given position of the $A_1$+$A_2$ system, the potential energy $U$ is taken from the nuclear matter self-energy with the corresponding density, isospin asymmetry, and energy. This is analogous to the local density approximation of Jeukenne, Lejeune, and Mahaux \cite{JLM}.

\begin{figure*}
\begin{center}
\includegraphics[scale=0.26]{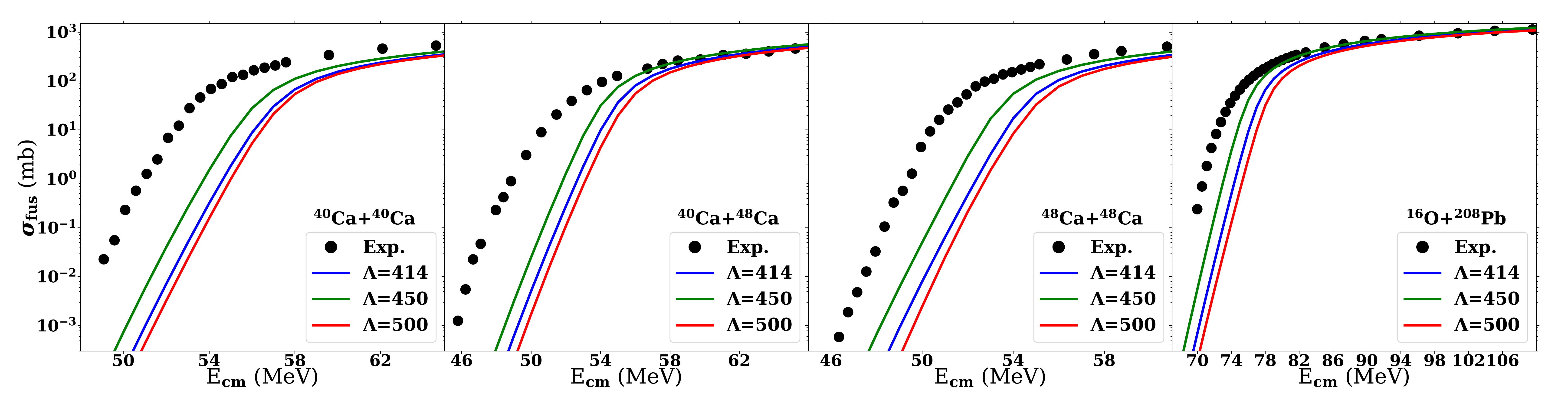}
\caption{Fusion cross sections for a variety of systems. Results from N$^3$LO chiral interactions with momentum space cutoffs of $\Lambda$ = 414, 450, 500 MeV are shown in blue, green, and red. The results do not include coupling effects. Experimental data from Refs. \cite{Jiang10,Montagnoli12,STEFANINI09,Morton99} are shown as black dots.} 
\label{FusionCrossSection}
\end{center}
\end{figure*} 

The two main ingredients needed for the present approach are the nucleon self-energy and the relevant nuclear density distributions. The nucleon self-energy is calculated to second order in many-body perturbation theory as a function of momentum in nuclear matter of given density and isospin asymmetry. The first- and second-order contributions to the nucleon self-energy are
\begin{equation}
\Sigma^{(1)}_{2N}(q;k_f)=\sum_{1} \langle \vec{q} \: \vec{h}_1 s s_1 t t_1 | \bar{V}_{2N}^{\rm eff} | \vec{q} \: \vec{h}_1 s s_1 t t_1 \rangle n_1 ,
\label{eq:2}
\end{equation}
\begin{eqnarray}
\label{sig2a}
&&\hspace{-.1in} \Sigma^{(2a)}_{2N}(q,\omega;k_f) \\ \nonumber
&&= \frac{1}{2} \sum_{123} \frac{|\langle \vec{p}_1 \vec{p}_3 s_1 s_3 t_1 t_3 | \bar{V}_{2N}^{\rm eff} | \vec{q} \vec{h}_2 s s_2 t t_2 \rangle|^2}{\omega+\epsilon_2-\epsilon_1-\epsilon_3+i\eta} \bar{n}_1 n_2 \bar{n}_3,
\end{eqnarray}
\begin{eqnarray}
\label{sig2b}
&&\hspace{-.1in} \Sigma^{(2b)}_{2N}(q,\omega;k_f) \\ \nonumber
&&= \frac{1}{2} \sum_{123} \frac{|\langle \vec{h}_1 \vec{h}_3 s_1 s_3 t_1 t_3 | \bar{V}_{2N}^{\rm eff} | \vec{q} \vec{p}_2 s s_2 t t_2 \rangle|^2}{\omega+\epsilon_2-\epsilon_1-\epsilon_3-i\eta}   n_1 \bar{n}_2 n_3,
\end{eqnarray}
where the first-order contribution is nonlocal, energy independent, and real, while the second-order contributions are nonlocal, energy dependent, and complex. For more details, see Refs. \cite{Whitehead20,Holt16}. The self-energy is calculated using a set of three chiral interactions to estimate the theoretical uncertainty of nuclear forces to nucleus-nucleus interactions and associated reaction observables. Nuclear densities are calculated in the Skyrme-Hartree-Fock framework using Skryme functionals that are tuned to reproduce microscopic nuclear matter equations of state calculated from the same set of chiral interactions used for the self-energy. See Ref. \cite{Lim17} for more details on how the Skyrme functionals are constructed. The optical potential is then calculated in the method described above using the densities and self-energies from each of the chiral interactions for a selection of $A_1$+$A_2$ systems at varying energies. At no point in the calculation are any phenomenological adjustments made to improve agreement with experimental data.

\emph{Results} - To assess the effectiveness of the approach outlined above, nucleus-nucleus interactions are computed for several cases where elastic scattering and fusion data are available for comparison of theoretical predictions with experiment. The nucleus-nucleus interactions produced in the present framework are substantially different from those produced in the commonly used double folding approach. As representative examples, the optical potentials for three different systems are shown in Fig. \ref{Potential}. The real part of the potentials are plotted with solid lines and the imaginary with dashed lines; the colors represent different chiral forces. The light grey region corresponds to densities of the $A_1$+$A_2$ system that exceed nuclear saturation density in the frozen density approximation. The inner dark grey region represents the part of the potential that does not affect elastic scattering. The size of this region depends on both the mass of each nucleus and the energy of the reaction; larger masses and lower energies produce a larger inner region of the potential that does not affect elastic scattering. The real term is attractive at large distances and becomes strongly repulsive in the interior. As with nucleon-nucleus optical potentials, the real term decreases in magnitude for increasing scattering energies while the opposite is true for the imaginary term. The imaginary term is absorptive at large distances and decreases in magnitude and changes sign in the interior with the exception of the $\Lambda=$414, 450 MeV potentials in the bottom plot of Fig. \ref{Potential} where the large scattering energy produces a very absorptive imaginary term. Note that due to the supersaturation densities resulting from the frozen density approximation, the imaginary potential can become positive and therefore unphysical. For these regions the imaginary term is set to zero. 

In Fig. \ref{CrossSection}, elastic scattering cross sections are shown for nucleus-nucleus systems ranging in mass from $^{16}$O+$^{28}$Si to $^{86}$Kr+$^{208}$Pb and ranging in projectile energy from $E_{\text{Lab}}$=6 MeV/N to $E_{\text{Lab}}$=94 MeV/N. The top row shows results at low energies ranging from $E_{\text{Lab}}$=6 MeV/N to $E_{\text{Lab}}$=20 MeV/N where collective excitations are relevant especially for smaller nuclei \cite{TREFZ84,Minomo16}. The agreement with experimental data is good for small and medium angles in all cases, but for large angles there is significant disagreement. These large angles correspond to scattering processes with larger momentum transfer that probe deeper into the nucleus-nucleus potential where the frozen density approximation is unreliable. The second row shows reactions at $E_{\text{Lab}} \sim$44 MeV/N for relatively large systems. The results in these cases agree very well with experimental data for each of the chiral interactions. Comparisons to the first three reactions in this row are also made in Ref. \cite{TREFZ85} where the authors carry out a similar calculation to the current work, although they use the Reid soft-core $N\!N$ interaction. Compared to Ref. \cite{TREFZ85} the results of the present work are in better agreement with data, even though the former includes surface excitations. Moving to higher energies, the third row shows reactions at $E_{\text{Lab}}$=94 MeV/N for systems ranging from medium to heavy mass. The results are in fair agreement with experiment, but are beginning to show the over-absorption characteristic of nuclear matter calculations. In the case of $^{16}$O+$^{28}$Si at $E_{\text{Lab}}$=94 MeV/N the $\Lambda=$500 MeV results overpredict for larger angles. This can be explained by noting that in Fig. \ref{Potential} the $\Lambda=$500 MeV imaginary terms tend to be less absorptive and approach zero at larger separation distances than the other two potentials.

In Fig. \ref{FusionCrossSection} fusion cross sections are shown for a range of systems. Fusion cross sections are calculated using only the real part of the nucleus-nucleus interaction in the code CCFULL \cite{HAGINO99}. At these energies of $\sim$2-3 MeV/N the imaginary part is small, however, it would be instructive to implement the imaginary term in future works. The fusion cross sections for calcium isotopes are substantially underpredicted at energies near the Coulomb barrier compared to experiment. This is partly due to the absence of coupling effects that are expected to enhance the cross section for energies near the barrier. However the agreement with experiment improves for energies above the barrier where the coupling effects become less significant and the form of the nuclear potential dominates. In the case of $^{16}$O+$^{208}$Pb fusion, the results show good agreement with experiment down to near-barrier energies below which the results then underpredict experiment. As with elastic scattering, the results for larger systems are in better agreement with experiment. It is important to note that while Pauli blocking is fully accounted for in calculating the self-energy, Pauli blocking between the two nuclei is not accounted for in the frozen density approximation. In Ref. \cite{Simenel17}, the authors follow a similar energy density approach for deriving a nucleus-nucleus interaction and show that accounting for Pauli effects can significantly improve agreement with the experimental fusion cross sections near the Coulomb barrier for the cases shown in Fig. \ref{FusionCrossSection}. Hence, the results of the current work may be improved if the nucleus-nucleus density is calculated in a way that accounts for Pauli effects and if couplings are included.

\emph{Summary} - To address the need for improved microscopic calculations of nuclear reaction models in the rare isotope beam era, a method for computing the nucleus-nucleus optical potential from nuclear matter calculations through an energy density functional has been carried out with a set of chiral interactions including two- and three-body forces. The method is successful at reproducing experimental results for medium mass and heavy systems at moderate energies $20 \lesssim E \lesssim 90$ MeV/N that are above the typical energies of collective excitations and below the over-absorption typical of nuclear matter derived optical potentials. The successful extension to lighter nuclei and lower energies may require the inclusion of surface excitations as shown in \cite{TREFZ84,Minomo16}. Additionally, improvements to the frozen density approximation will be investigated in future works. If the nucleus-nucleus optical potentials constructed using the present method vary smoothly enough with mass and energy, and a suitable parametric form is found to describe the potential, a microscopic global model may be achievable.

\begin{acknowledgments}
The author thanks J. W. Holt for providing nuclear matter calculations. The author also acknowledges F. M. Nunes, K. Godbey, and C. Hebborn for informative discussions. This work was supported by the U.S. Department of Energy (Office of Science, Nuclear Physics) under grant DE-SC0021422.
\end{acknowledgments}


%

\end{document}